\documentclass[12pt]{iopart}

\usepackage{iopams}  
\usepackage{graphicx}
\expandafter\let\csname equation*\endcsname\relax
\expandafter\let\csname endequation*\endcsname\relax
\usepackage{amsmath}
\usepackage{gensymb}
\usepackage{color} 
\usepackage{hyperref}
\hypersetup{
colorlinks=true,final=true,
        linkcolor=blue,
        citecolor=blue,
        filecolor=blue,
        urlcolor=blue,
}

\begin{document}

\title{Cell damage and mitigation in Swiss albino mice: experiment and modelling}

\author{Urmimala Dey$^*$}
\address{Centre for Theoretical Studies, Indian Institute of Technology Kharagpur, Kharagpur 721302, India}
\author{Archisman Ghosh}
\address{Department of Environmental Carcinogenesis and Toxicology, Chittaranjan National Cancer Institute, 37 S. P. Mukherjee Road, Kolkata
700026, India}
\author{Syed Abbas}
\address{School of Basic Sciences, Indian Institute of Technology, Mandi 751005, India}
\author{A Taraphder}
\address{Department of Physics, Indian Institute of Technology Kharagpur, Kharagpur 721302, India}
\address{Centre for Theoretical Studies, Indian Institute of Technology Kharagpur, Kharagpur 721302, India}
\address{School of Basic Sciences, Indian Institute of Technology, Mandi 751005, India}
\author{Madhumita Roy}
\address{Department of Environmental Carcinogenesis and Toxicology, Chittaranjan National Cancer Institute, 37 S. P. Mukherjee Road, Kolkata 700026, India}

\ead{$^*$urmimaladey@iitkgp.ac.in}
\vspace{10pt}
\centerline{(\today)}
\vspace{0.5cm}

\begin{abstract}
Chronic exposure to inorganic arsenic is a potential cause of carcinogenesis. It elicits its potential by generation of ROS, leading to DNA, protein and lipid damage. Therefore, the deleterious effect of arsenic can be mitigated by quenching ROS using antioxidants.  There is a homology between the protein coding regions of mice and human. Effect of these alterations in human can be mimicked in mice. Therefore to understand the underlying mechanism of arsenic toxicity and its amelioration by black tea, studies have been conducted in mice model. Long term exposure to iAs leads to tumour growth, which has been found to be alleviated by black tea. Observations reveal that black tea has two salutary effects on the growth of tumour: the rate of growth of damaged cells was appreciably reduced and an early saturation of the level of damage is achieved. To take the experimental findings further, the experimental data have been modelled with simple dynamical equations. The curves obtained from \textit{in vivo} studies have been fitted with the data obtained from the model. The corresponding steady states and their stabilities are analyzed.
\end{abstract}

\section{Introduction}
\noindent Arsenic, the 20th most abundant element in the earth's crust~\cite{Herath_2016} is found in nature in two oxidation states i.e. +3 (arsenite) and +5 (arsenate). It is a metalloid, found in both organic and inorganic forms. Arsenic is released into the ground water from rocks containing copper or lead, either naturally or through intensive anthropogenic activities like mining etc~\cite{Victor_2011}. Humans may be exposed to As primarily from air, food and drinking water~\cite{Biswas_2010}. Chronic exposure to inorganic arsenic (iAs) has a strong correlation with many health hazards. More than 200 million people residing in 40 countries around the world are exposed to dangerous levels of inorganic As (iAs) through ground water, the level being much higher than the permissible limit of 10 $\mu$g/l~\cite{Khairul_2017}. In the Indian subcontinent, at several places, arsenic levels of water range from 50-3200 mg/l~\cite{Biswas_2010}.

\noindent Exposure to iAs leads to development of several diseases including cancer. Excessive generation of reactive oxygen species (ROS) by iAs is considered to be the main reason behind the imbalance and disturbed homeostasis of individuals~\cite{Sinha_2010}. ROS causes damage to DNA, protein and lipid and modulates expression~\cite{Ramos_2009} of several genes including those implicated in cancer. Chronic arsenic exposure has also been reported to decrease the activities of antioxidant enzymes which are required to counter the effect of free radicals due to ROS generation~\cite{Sinha_2010}.

\noindent Carcinogenesis is a multistep process involving initiation, promotion, progression and finally metastasis. In order to check the deleterious effect of iAs, it is preferable to halt the phenomenon of carcinogenesis at the initial stages itself. Chemoprevention is the use of natural, synthetic, or biologic agents in order to reverse, suppress, or prevent the progression of cancer~\cite{Tsao_2004}. Chemopreventive strategies to mitigate the effect of iAs-induced toxicity using natural compounds is the need of the hour. Tea, the most preferred beverage is a good antioxidant and has anticancer properties~\cite{Sinha_2005}. Owing to its antioxidant properties, black tea can counter the deleterious effects of ROS generated due to iAs. To elucidate the role of black tea in the prevention of iAs-induced carcinogenesis, studies have been initiated in Swiss Albino mice.  Black tea can efficiently quench generation of ROS, diminish DNA damage and lipid peroxidation, eliciting its anti-carcinogenic potential.
\section{Materials and methods}
\subsection{Chemicals}
\noindent Preparation of Tea Extracts \\
\noindent 1.25 gm of Assam tea and 1.25 gm of Darjeeling tea was mixed together and brewed in 100 ml of freshly boiled water for 5 minutes. The liquor was then lyophilized using a SCANVAC lyophilizer and stored. The lyophilized powder was then weighed and reconstituted in water and administered to mice by gavage. The catechin and theaflavin contents of the particular brand of green tea and black tea has been analysed by HPLC. Dose was determined by epidemiological evidences and previous publications from our lab~\cite{Sinha_2010}. Accordingly 100 $\mu$l of tea extract was administered to the mice.
\begin{figure}[h]
\centering
\includegraphics[scale=.40]{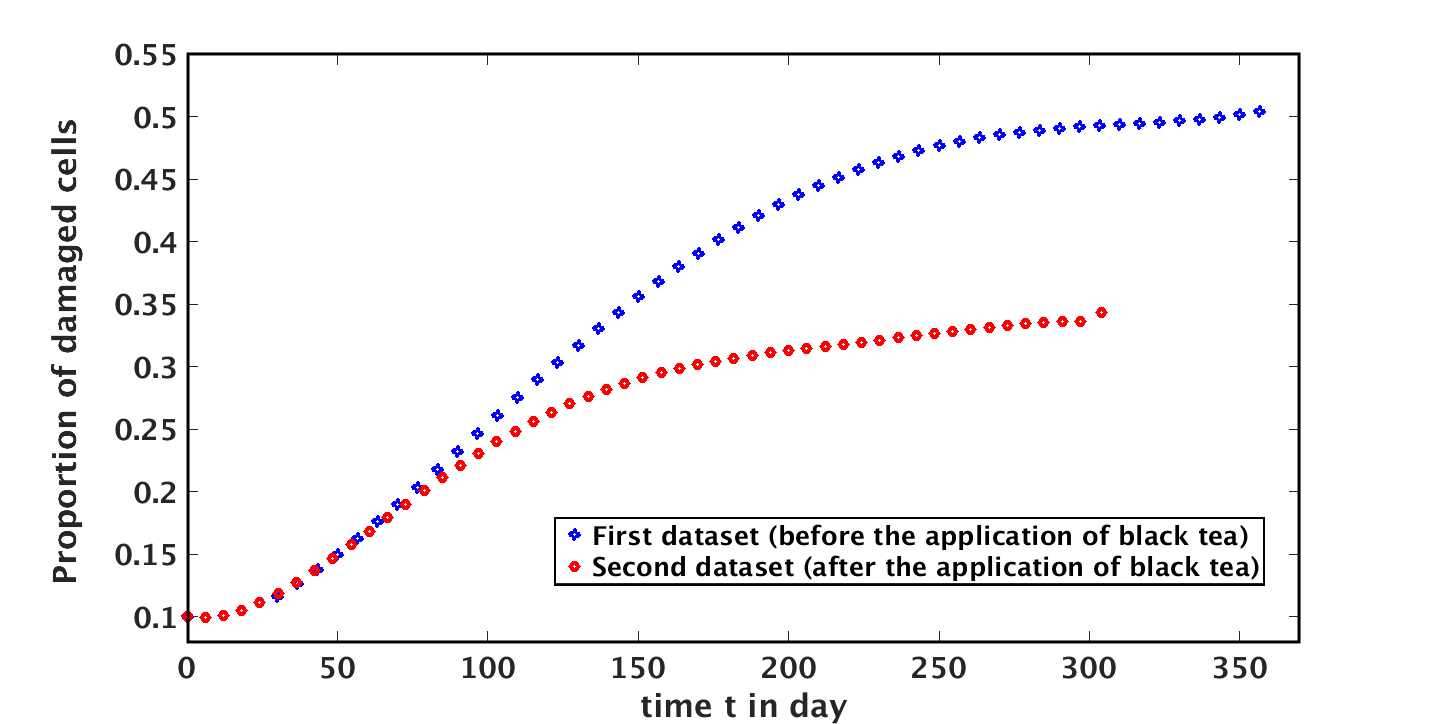}
\caption{Experimental datasets for the proportion of cell damage due to iAs (upper curve). The lower curve is for the group that received both iAs and black tea. Note the amelioration due to black tea, in limiting the extent of damage and causing an early saturation.}
\label{exp_plot}
\end{figure}
\subsection{Maintenance of animals}
\noindent Normal, male, 4-5 week old Swiss albino mice (Mus Musculus) were selected from an inbred colony maintained in the animal house of the Institute (IAEC – 1774/MR-3/2017/9). During the course of study the animals were fed on synthetic pellets specified for mice-feed and had access to water ad libitum. They were maintained under standard conditions with a 12 hour light and dark period and controlled room temperature (22$\pm$2\degree C). The animals were sacrificed after euthanasia with thiopentone sodium in overdose (100 mg/kg body wt.).
\subsection{Treatment}
\noindent The mice were divided in four groups namely I, II, III and IV.
\noindent Group I: Control group, where the mice were fed with normal food and tap water.
\noindent Group II: Mice were fed with black tea by gavage.
\noindent Group III: Mice were given arsenic water (500 $\mu$g/l) for a period of 360 days.
\noindent Group IV: Black tea (100 mg/ml, thrice daily) was administered to the mice by gavage along with arsenic water (500 $\mu$g/ml) for the same period of duration. On sacrificing the mice at different time intervals, blood and tissue samples were collected and stored properly to perform the experiments. Two sets of data estimating the damaged cells with time (up to about 360 days) were generated : one is only with iAs and the other, with both iAs and black tea (for this group, the data were available for about 330 days). No dataset for Group II is presented because no toxicity is found due to the application of black tea. There were already some cell damage in the mice and therefore the damage cell count does not start from zero at zero time. Fig.~\ref{exp_plot} shows both these data: the application of black tea clearly brings down the cell damage and leads to early saturation. 
\subsection{Estimation of ROS generation}
Generation of Reactive Oxygen species was estimated in the blood of the mice according to Balasubramanyam \textit{et al.}~\cite{Balasubramanyam_2003} with slight modifications. Leukocytes were isolated from blood
using solution A (NH$_4$Cl dissolved in TRIS at pH 7.2) and Solution B (meso-inositol), then suspended in HBS (Hepes Buffered Saline pH 7.4). 10 mM 2$'$,7$'$ dichloro-dihydro-fluorescein (DCFH-DA) was added and incubated for 45 minutes in the dark. The fluorescent intensity of the DCF was recorded at 530 nm .
\subsection{Single Cell Gel Electrophoresis}
DNA damage was estimated by the Single Cell Gel Electrophoresis (SCGE) or Comet assay~\cite{Singh_1988}, using leukocytes isolated from blood. 20 $\mu$l of leukocytes suspended in 0.6\% Low melting agarose (LMA) and were layered over frosted slide pre-coated with 0.75\% Normal melting agarose (NMA). Cells were subjected to lysis by immersing them in ‘lysis buffer’ (2.5M NaCl, 0.1M Na$_2$EDTA, 10mM TRIS, 0.3M NaOH, 10\%DMSO and 1\% TritonX; pH 10) for overnight at 4\degree C. Slides were then presoaked in electrophoresis buffer (10mM NaOH, 0.2 m Na$_2$EDTA; pH$>$13) for 20 minutes and subsequently subjected to electrophoresis for 25 minutes at 15V, 220mA. After staining with ethidium bromide (EtBr), slides were viewed under a fluorescent microscope.
\subsection{Lipid peroxidation assay}
Damage to the cell membrane due to ROS generation was measured using the lipid peroxidation assay following the principle of Okhawa \textit{et al.}~\cite{Ohkawa_1979}, with slight modifications. The liver tissue of the sacrificed animals were collected and homogenized. 10\% SDS, 20\% acetic acid and 0.8\% TBA was added to the tissue homogenate. It was then placed in boiling water bath for 1 hour and then immediately transferred to ice for 10 minutes. The samples were then centrifuged at 2500 rpm for 10 minutes. The supernatant was collected and absorbance was measured at 535 nm. Lipid peroxidation was calculated as the number of moles of MDA (Malondialdehyde) generated.

\section{Model for growth of cell damage and mitigation}
\noindent The dynamics of the effects of iAs and black tea on the colony of Swiss albino mice are modelled by a set of simple mathematical equations. As mentioned above, we have two datasets that represent the growth of damaged cells in the groups of mice studied, one with only iAs (at 500 $\mu g$/l) and the other with tea (100 mg/ml, thrice daily), acting, as it does, as a mitigating agent and iAs with the same dose as above.  The extent of cell damage is proportional to the generation of reactive oxygen species (ROS).
\noindent First we ignore the effect of immunity and try to understand the effects of iAs and tea only. The effects of immune cell response will be brought on in the next section.  Our aim in this and the following section is to try to model actual data that are obtained from our experiment using standard mathematical models. There are excellent review articles on such models and their merits and demerits (see, for example, Altrock, \textit{et al.}~\cite{Altrock_2015} and Pillis, \textit{et al.},~\cite{Pillis_2001}). We do not, therefore, digress on these details here.

\noindent The dynamics of the damaged cells may be represented by the usual growth equation:
\begin{equation}
\frac{dn}{dt} = rn(1-\frac{n}{K})
\label{eq1}
\end{equation}
$N$ in the above equation is a measure of the damaged cells as interpreted from ROS data. We define $n=\frac{N}{N_0}$ where $N_0$ is the total number of cells at the beginning. Note that one does not need to know the value of either $N$ or $N_0$, only the ratio (percentage) of damaged cells, represented by $n$, is needed for the model. There are two unknown constants above, $r$ and $K = B/N_0$, where $B$ is the well-known `carrying capacity'. \\
Now iAs is administered at a constant rate daily from outside and is incorporated via the term $\alpha A n$ in the above equation while the mitigating effects of tea is introduced via the term $\beta T n$~\cite{Liu_2017}. Then the overall equation for the cell growth is
\begin{equation}
\frac{dn}{dt} = rn(1-\frac{n}{K}) + \alpha A n - \beta T n
\label{eq2}
\end{equation}
The constants $\alpha$ and $\beta$ are unknown, to be fitted from the data. The equations governing the concentration of iAs and tea, $A$ and $T$ respectively, are similar~\cite{Liu_2017}. In principle, there could be additional effects, but in the minimal model we consider the effects of iAs and tea require the two terms shown below:
\begin{equation}
\frac{dA}{dt} = A_0 - \gamma_A A
\label{eq3}
\end{equation}
\begin{equation}
\frac{dT}{dt} = T_0 - \gamma_T T
\label{eq4}
\end{equation}
Here $A_0$ and $T_0$ are the constant rate of administration of iAs and tea externally. Here again, other effects of iAs and tea on the body could be added, which we will consider in the future. One effect that is usually considered important, but we neglect here (discussed in the next section), is the immune-surveillance. Usually, $A_0$ and $T_0$, the so-called dosage, are known in terms of mg (or ml)/day. However, to incorporate them in the Eq.~(\ref{eq2}) above, one needs to know the concentration thereof at the cellular level, which, at this moment, is not available to us. Therefore we leave them to fitting. For the dataset with iAs only, the equation for tea is obviously redundant. \\
The expulsion of iAs and tea from the body is also accounted for via decay rates (rough estimates of half-lives of elimination for these are available), $\gamma_A$ and $\gamma_T$, from the body. Note that different ingredients decay differently, depending also on the organ considered. So some approximate values are chosen from literature~\cite{Hughes_2003}, though a considerable degree of latitude could not be ruled out.\\
Solving Eq.~(\ref{eq3}) and Eq.~(\ref{eq4})
\begin{equation}
A(t) = \frac{A_0}{\gamma_A}(1-e^{-\gamma_A t}) + A_i e^{-\gamma_A t}
\label{eq5}
\end{equation}
\begin{equation}
T(t) = \frac{T_0}{\gamma_T}(1-e^{-\gamma_T t}) + T_i e^{-\gamma_T t}
\label{eq6}
\end{equation}
where, $A_i = A(0)$ and $T_i = T(0)$.
Substituting the expressions of $A(t)$ and $T(t)$ in Eq.~(\ref{eq2}) :
\begin{equation}
\frac{dn}{dt} = rn(1-\frac{n}{K}) + m n (1 - e^{-\gamma_A t}) + p n e^{-\gamma_A t} - q n (1 - e^{-\gamma_T t}) - s n e^{-\gamma_T t}
\label{eq7}
\end{equation}
where, $m$ = $\frac{\alpha A_0}{\gamma_A}$, $p$ = $\alpha A_i$, $q$ = $\frac{\beta T_0}{\gamma_T}$ and $s$ = $\beta T_i$.
\begin{figure}[h]
\centering
\includegraphics[scale=.42]{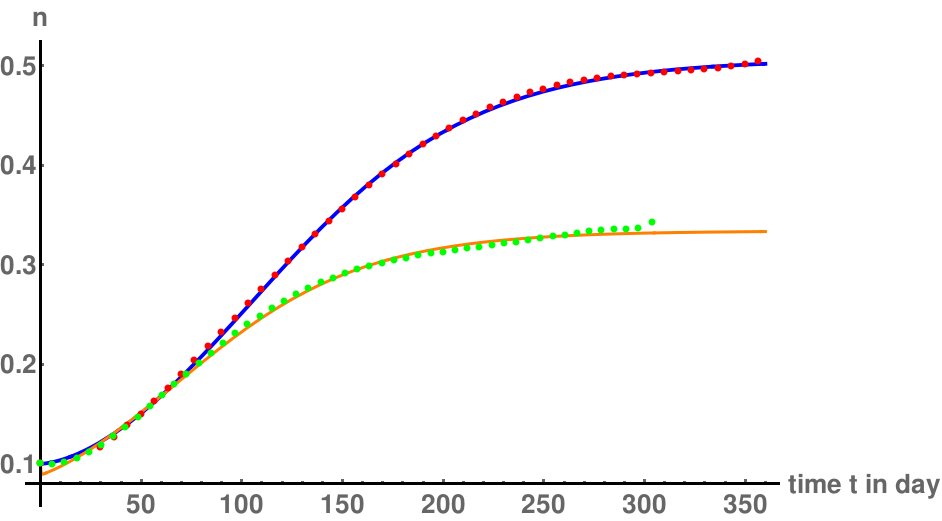}
\caption{Curves fitted to the given datasets. Here, red and green dots represent experimental data and the blue and orange solid lines represent the fitted curves. }
\label{fit}
\end{figure}
It is safe to assume that there is no prior presence of iAs and tea in the mice. Normal tap water which is given to the mice is tested for the presence of iAs (by atomic absorption spectroscopy (AAS)). The tap water was found to be free of the metalloid. Although the presence of iAs at a very small level in the body cannot be ruled out, but that is still negligible for the model. The initial values of iAs and tea at time $t=0$ i.e., $A_i$, $T_i$ = $0$. This implies $p,\,s$ = 0. In our case, $\gamma_A$ = 0.04 day$^{-1}$ and $\gamma_T$ = 4 day$^{-1}$. We reiterate that these values are merely approximations. Such rates depend on a large number of factors, primarily the organ being considered, the method of administration and the measurement protocol. The values used here are similar to Hughes \textit{et al.}~\cite{Hughes_2003} Solving and fitting Eq.~(\ref{eq7}) numerically to the given datasets, we obtain the values of the unknown coefficients as : $r$ = 0.00524, $K$ = 0.14706, $m$ = 0.01281 for the first dataset and $r$ = 0.01380, $K$ = 0.21644, $m$ = 0.00792, $q$ = 0.00041 for the second dataset. Note that for the first dataset, the effect of black tea is not included. So, $q$ is not relevant for the first dataset. \\
Fig.~\ref{fit} shows the curves fitted to the given datasets, red and green dots represent experimental data. The fitting appears to be excellent for both the curves. Note that although iAs has a much longer half-life compared to black tea, the competition between the damage by iAs and mitigation by tea is clear from the data and the model. The saturation value and time for the curve with tea and iAs is considerably lower than the upper curve with iAs alone. This is a redeeming feature from the experiment and the model that the effect of tea is quite appreciable. \\
The fact that curves for iAs alone and iAs and tea together could both be fitted with similar models gives confidence that the model could be used for further predictions. One such prediction is that the damage appears to saturate in both cases, earlier with tea. Such saturation would imply that the damage could be limited by the use of an external mitigant like tea. We have neglected the effects of immunogenicity above, which will abet early saturation. Although in the present paper we took the actual data and modelled them, which makes modelling much more relevant, it is important to underline the remit of data-fitting by mathematical  equations : such exercises should be carried out with due caution. In the absence of cellular level concentration of iAs and black tea, and the inherent uncertainty and fluctuation in their values at the site, the results of the model should be taken as indicative only. In addition, as we show below, the effect of immune-cell response dramatically alters the model and make it closer to reality~\cite{Pillis_2003}.
\section{Growth model including the effect of immune cells}
\begin{figure}[h]
\centering
\includegraphics[scale=.42]{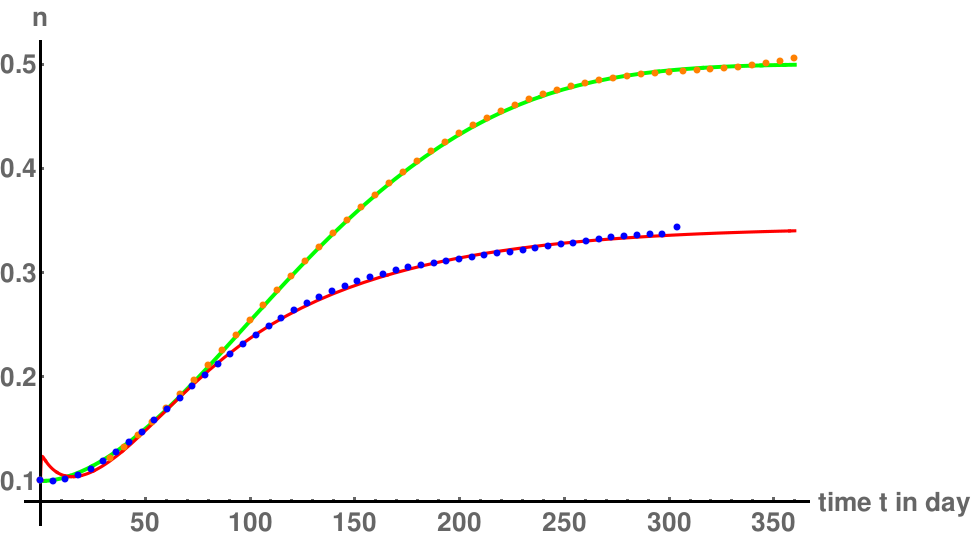}
\caption{Curves fitted to the given datasets. Here, orange and blue dots represent experimental data and the green and red solid lines represent the fitted curves. }
\label{fit2}
\end{figure}

Let $I(t)$ denote the number of immune cells at any instance of time t with a constant influx rate of $I_0$ and if the death rate of the immune cells be $\gamma_I$, the overall dynamics of the system may be represented by the following equations :
\begin{equation}
\frac{dn}{dt} = rn(1-\frac{n}{K}) + \alpha A n - \beta n T - \epsilon n I
\label{eq8}
\end{equation}
\begin{equation}
\frac{dA}{dt} = A_0 - \gamma_A A
\label{eq9}
\end{equation}
\begin{equation}
\frac{dT}{dt} = T_0 - \gamma_T T
\label{eq10}
\end{equation}
\begin{equation}
\frac{dI}{dt} = I_0 + \frac{\rho I n}{\alpha_1 + n} -c_1 I n - \gamma_I I
\label{eq11}
\end{equation}
The decay rates of iAs and black tea ($\gamma_A$ and $\gamma_T$) are given in the previous section. We assume the constant influx rate of the immune cells $I_0$ = 0.3 and the death rate of the immune cells $\gamma_I$ = 0.0208 day$^{-1}$~\cite{Pillis_2007, Kuznetsov_1994}. Eqs.~(\ref{eq9}) and (\ref{eq10}) can be solved analytically as given in Eqs.~(\ref{eq5}) and \ref{eq6}. Substituting the expressions of $A(t)$ and $T(t)$ in Eq.~(\ref{eq8}) as before, we obtain
\begin{equation}
\frac{dn}{dt} = rn(1-\frac{n}{K}) + m n (1 - e^{-\gamma_A t}) + p n e^{-\gamma_A t} - q n (1 - e^{-\gamma_T t}) - s n e^{-\gamma_T t} - \epsilon n I
\label{eq12}
\end{equation}
where, the unknown coefficients $m$ (= $\frac{\alpha A_0}{\gamma_A}$), $p$ (= $\alpha A_i$), $q$ (= $\frac{\beta T_0}{\gamma_T}$), $s$ (= $\beta T_i$), $\epsilon$, $\rho$, $\alpha_1$ and $c_1$ are to be fitted from the experimental data. We consider that initially at $t$ = 0, no iAs, tea and immune cells are present. This implies $A_i$, $T_i$ = 0. Therefore, the coefficients $p$, $s$ = 0. \\
Numerically solving Eqs.~(\ref{eq11}) and (\ref{eq12}) and after fitting to the experimental datasets, we obtained the values of the unknown coefficients as : $r$ = 0.00173, $K$ = 0.03166, $m$ = 0.02802, $\epsilon$ = 0.00055, $\rho$ = 0.95103, $\alpha_1$ = 1.30187, $c_1$ = 0.63433 for the first dataset and $r$ = 0.60516, $K$ = 8.97523, $m$ = 0.08006, $q$ = 0.61760, $\epsilon$ = 0.00379, $\rho$ = 0.02733, $\alpha_1$ = 2.15877, $c_1$ = 0.02718 for the second dataset. For the first dataset, the coefficients are calculated without considering the effect of black tea. The fitted curves are given in Fig.~\ref{fit2} along with the experimental datasets, which are shown by the orange and blue dots. The existence and uniqueness of the solutions are discussed in the Appendix.

\subsection{Equilibrium points}
\subsubsection{Tea-free equilibria}
\begin{figure}[h]
\centering
\includegraphics[scale=.42]{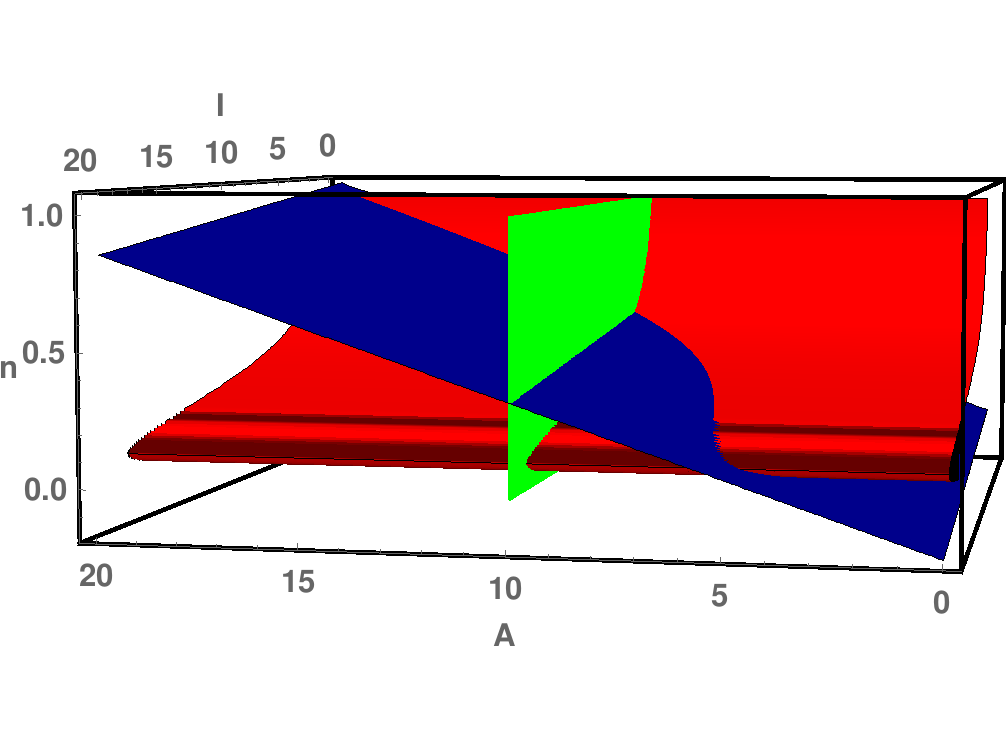}
\caption{The three null-surfaces given by Eqs.~(\ref{eq16}$-$\ref{eq18}) using the values of the coefficients obtained from the fit. Here, we have taken $A_0$ = 0.4. The equilibrium points ($n^*$, $A^*$, $T^*$, \,$I^*$) appear at the intersections of these surfaces. }
\label{no_tea}
\end{figure}

In order to study the equilibrium solutions, we first consider the case without tea. We look for the steady state solutions:
\begin{equation*}
\dot{n} = 0
\end{equation*}
\begin{equation*}
\dot{A} = 0
\end{equation*}
\begin{equation*}
\dot{I} = 0
\end{equation*}
For experimental relevance, we look for the non-zero solution for $n$ only. The three Eqs. above generate three null surfaces, the intersection of which give the equilibrium point ($n^*$,\, $A^*$, \, $I^*$).  The non-trivial solutions are :
\begin{equation}
n^* = K \bigg(1 + \frac{\alpha A^*}{r} - \frac{\epsilon I^*}{r}\bigg)
\label{eq16}
\end{equation}
\begin{equation}
A^* = \frac{A_0}{\gamma_A}
\label{eq17}
\end{equation}
\begin{equation}
I^* = \frac{I_0(\alpha_1 + n^*)}{c_1 \alpha_1 n^* + c_1 {n^*}^2 + \gamma_I \alpha_1 + \gamma_I n^* - \rho n^*}
\label{eq18}
\end{equation}
\noindent Although the trivial equilibrium (no-tumour) $n^{*}=0$ is studied for the sake of completeness by other authors~\cite{Pillis_2003}, it does not correspond to any relevant experimental outcome. Therefore we neglect the trivial point. In Fig.~\ref{no_tea}, we plot the three null-surfaces given by Eqs.~(\ref{eq16}$-$\ref{eq18}) using the values of the coefficients, as obtained from the fit. Solving Eqs.~(\ref{eq16}) and (\ref{eq18}) numerically and using the obtained values of the coefficients, we find that equilibrium occurs at the point ($A^*$,  $I^*$, $n^*$) = (10, 4.010730, 0.502892), where we consider the constant influx rate of iAs is fixed at $A_0$ = 0.4. Once the non-trivial points of equilibrium are obtained, we study their linear stability. In order to ensure the positivity of equilibria, we need $r + \alpha A^*> \epsilon I^*$ and
$\rho>c_1\alpha_1+\gamma_I.$ The other condition we obtain by observing the roots of
$c_1 {n^*}^2 +(c_1 \alpha_1+\gamma_I- \rho) n^*+ \gamma_I \alpha_1=c_1 {n^*}^2 +b n^*+ \gamma_I \alpha_1=(n^*-a_1)(n^*-a_2),$ where
$a_{1,2}=\frac{1}{2c_1}(-b \pm \sqrt{b^2-4c_1\gamma_I \alpha_1}).$

\subsubsection{Stability analysis}
We construct the Jacobian matrix:
\begin{equation*}
\begin{split}
J &= \begin{pmatrix}
    \alpha A^* - \epsilon I^* + r - \frac{2r}{K} n^* & \alpha n^* & -\epsilon n^* \\
    0 & -\gamma_A & 0\\
   \frac{\rho I^*}{\alpha_1 + n^*} - \frac{\rho I^* n^*}{{(\alpha_1 + n^*)}^2} - c_1 I^*& 0 & \frac{\rho n^*}{\alpha_1 + n^*} - c_1 n^* - \gamma_I
\end{pmatrix} \\
&= \begin{pmatrix}
    L^* & \alpha n^* & -\epsilon n^* \\
    0 & -\gamma_A & 0\\
    M^*& 0 & P^*
\end{pmatrix}
\end{split}
\end{equation*}
where, $L^*$ = $\alpha A^* - \epsilon I^* + r - \frac{2r}{K} n^*$, $M^*$ = $\frac{\rho I^*}{\alpha_1 + n^*} - \frac{\rho I^* n^*}{{(\alpha_1 + n^*)}^2} - c_1 I^*$ and $P^*$ = $\frac{\rho n^*}{\alpha_1 + n^*} - c_1 n^* - \gamma_I$ \\
The three eigenvalues of $J$ are given by:
\begin{equation}
\lambda_1 = -\gamma_A < 0
\label{eq19}
\end{equation}
\begin{equation}
\lambda_{\pm} = \frac{1}{2} \bigg[(L^*+P^*) \pm \sqrt{{(L^*-P^*)}^2 - 4\epsilon n^* Q^*}\bigg]
\label{eq20}
\end{equation}
For stability, eigenvalues must be negative. So, we have to find the region of negative eigenvalues along the $n-I-A$ axes.
Substituting the values of $n^*$, $A^*$ and $I^*$ in Eq.~(\ref{eq20}), we get $\lambda_{\pm} < 0$. Since $\lambda_1$ is already negative, it corresponds to stable equilibrium.
\subsubsection{Equilibrium with Tea}
Following the same procedure, we solve the equations:
\begin{equation*}
\dot{n} = 0
\end{equation*}
\begin{equation*}
\dot{A} = 0
\end{equation*}
\begin{equation*}
\dot{T} = 0
\end{equation*}
\begin{equation*}
\dot{I} = 0
\end{equation*}
The non-zero solutions are
\begin{equation}
n^* = K \bigg(1 + \frac{\alpha A^*}{r} - \frac{\beta T^*}{r} - \frac{\epsilon I^*}{r}\bigg)
\label{eq21}
\end{equation}
\begin{equation}
A^* = \frac{A_0}{\gamma_A}
\label{eq22}
\end{equation}
\begin{equation}
T^* = \frac{T_0}{\gamma_T}
\label{eq23}
\end{equation}
\begin{equation}
I^* = \frac{I_0(\alpha_1 + n^*)}{c_1 \alpha_1 n^* + c_1 {n^*}^2 + \gamma_I \alpha_1 + \gamma_I n^* - \rho n^*}
\label{eq24}
\end{equation}
To ensure the positivity of equilibria, we need $r + \alpha A^*> \beta T^*+\epsilon I^*$ and
$\rho>c_1\alpha_1+\gamma_I.$
\begin{figure}[h]
\centering
\includegraphics[scale=.28]{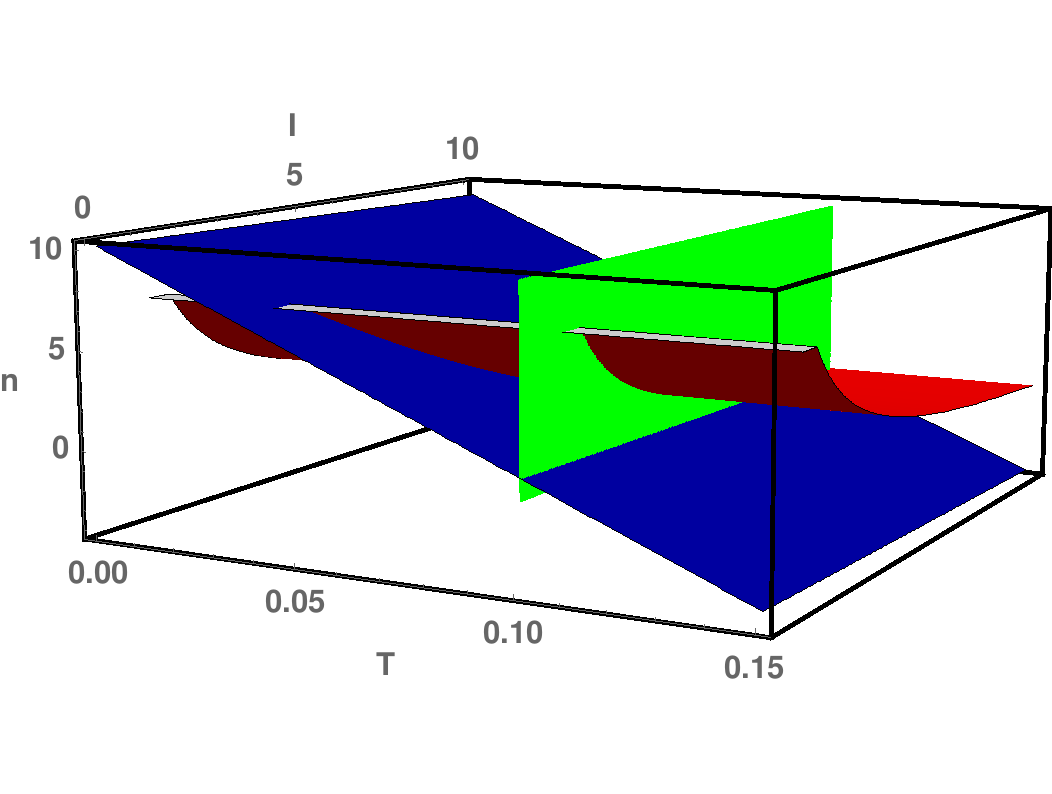}
\includegraphics[scale=.28]{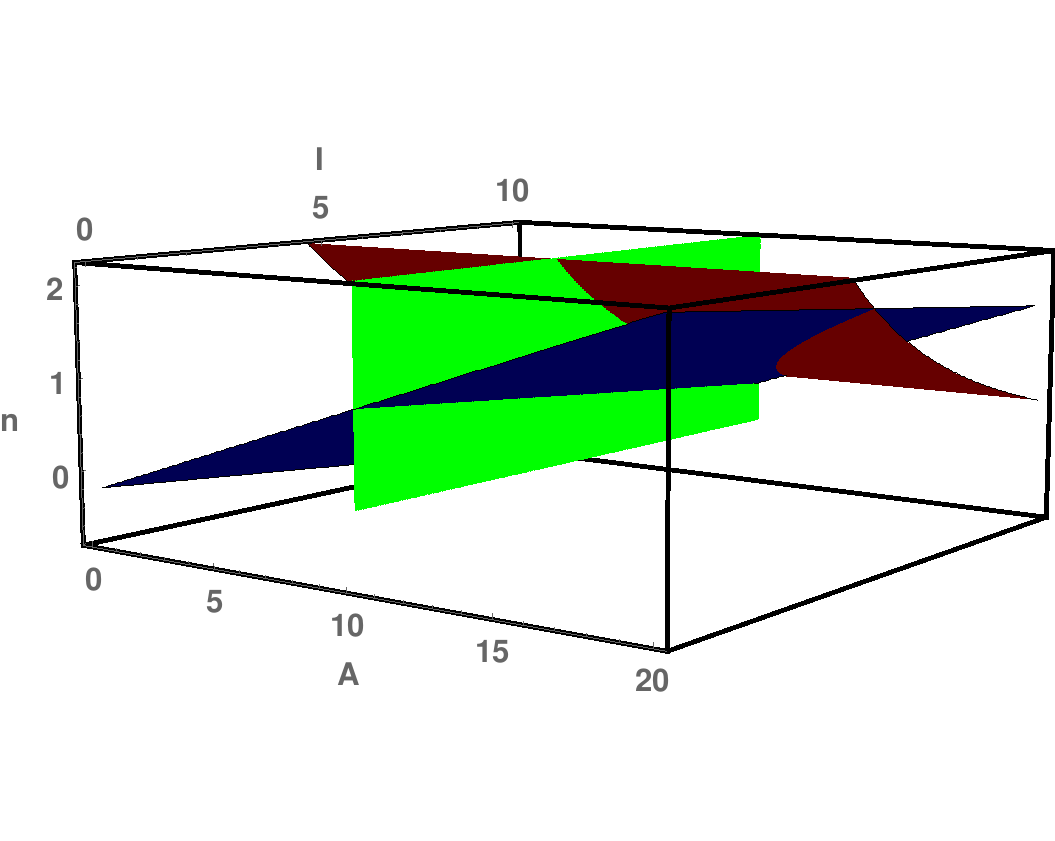}
\caption{The three null-surfaces given by Eqs.~(\ref{eq21}$-$\ref{eq24}) using the values of the coefficients, as obtained from the fit. Here, we have taken $A_0$ = 0.4 and $T_0$ = 0.4. The upper one is plotted keeping $A$ fixed at $A$ = $A^* = 10$ and in the lower one, the null-surfaces are plotted keeping $T$ fixed at $T$ = $T^* = 0.1$. The equilibrium points ($T^*$, $I^*$, $n^*$) and ($A^*$, $I^*$, $n^*$) occur at the intersections of these surfaces. }
\label{with_tea}
\end{figure}
The equilibrium points ($T^*$, $I^*$, $n^*$) and ($A^*$, $I^*$, $n^*$) are obtained from the upper and lower figures in Fig. \ref{with_tea}. They occur at the intersections of these surfaces as before.
\subsubsection{Stability analysis}
We write the 4$\times$4 Jacobian Matrix now
\begin{equation*}
J = \begin{pmatrix}
    L^* & \alpha n^* & -\beta n^* & -\epsilon n^* \\
    0 & -\gamma_A & 0 & 0\\
    0 & 0 & -\gamma_T & 0\\
    M^*& 0 & 0 & P^*
\end{pmatrix}
\end{equation*}
where, $L^*$ = $\alpha A^* - \beta T^* - \epsilon I^* + r - \frac{2r}{K} n^*$, $M^*$ = $\frac{\rho I^*}{\alpha_1 + n^*} - \frac{\rho I^* n^*}{{(\alpha_1 + n^*)}^2} - c_1 I^*$ and $P^*$ = $\frac{\rho n^*}{\alpha_1 + n^*} - c_1 n^* - \gamma_I$. The two eigenvalues of the Jacobian, $\lambda_1 = -\gamma_A$ and $\lambda_2 = -\gamma_T$ are both negative. The other two eigenvalues are
\begin{equation}
\lambda_{\pm} = \frac{1}{2} \bigg[(L^* + P^*) \pm \sqrt{{(L^* - P^*)}^2 - 4\epsilon n^* M^*}\bigg]
\label{eq25}
\end{equation}
Again we have to look for the regions where $\lambda_{\pm} < 0$. In Fig.~\ref{with_tea}, we show the three null-surfaces given by Eqs.~(\ref{eq21}$-$\ref{eq24}) using the values of the coefficients, as obtained from the fit; the upper one is plotted keeping $A$ fixed at $A^* = 10$. In the lower one, the null-surfaces are plotted keeping $T$ fixed at $T^* = 0.1$. The equilibrium points ($T^*$, $I^*$, $n^*$) and ($A^*$, $I^*$, $n^*$) are the intersections of these surfaces. Our calculations show that equilibrium occurs at the point ($A^*$, $T^*$, $I^*$, $n^*$) = (10, 0.1, 11.48990, 0.35624), where the constant influx rates of iAs and black tea are kept fixed at $A_0$, $T_0$ = 0.4. When we substitute these values of $n^*$, $A^*$, $T^*$ and $I^*$ in Eq.~(\ref{eq25}), we find that $\lambda_{\pm} < 0$. Since the other two eigenvalues are already negative, this implies that the values of the coefficients obtained from the fit actually lead to a stable equilibrium.

\subsection{Model with delay}
The immune response in any living organism is usually a delayed process and therefore it is natural to assume a delay in $I(t)$ in the functional response term. In this situation, the overall dynamics of the system may be represented by the following equations:
\begin{eqnarray}
\frac{dn(t)}{dt} &=& rn(t)(1-\frac{n(t)}{K}) + \alpha A(t) n(t) - \beta n(t) T(t) - \epsilon n(t) I(t) \nonumber \\
\frac{dA(t)}{dt} &=& A_0 - \gamma_A A(t) \nonumber \\
\frac{dT(t)}{dt} & = & T_0 - \gamma_T T(t) \nonumber \\
\frac{dI(t)}{dt} &=& I_0 + \frac{\rho I(t-\tau) n(t)}{\alpha_1 + n(t)} -c_1 I(t) n(t) - \gamma_I I(t),
\label{dm}
\end{eqnarray}
with history condition $n(0)=n_0, A(0)=A_0, T(0)=T_0, I(t)=\phi(t)$ for $t \in [-\tau,0].$

\subsubsection{Stability of Equilibrium without Tea}
Now we discuss the stability of non-trivial equilibrium solutions. It is to note that the set of equilibria are the same for this system as with Eqs.~(\ref{eq8})$-$(\ref{eq11}). First, we analyze tea free system.

\noindent We construct the Jacobian matrix :
\begin{equation*}
\begin{split}
\Delta(\lambda, \tau) &= det \begin{pmatrix}
    (\alpha A^* - \epsilon I^* + r - \frac{2r}{K} n^*)-\lambda & \alpha n^* & -\epsilon n^* \\
    0 & -\gamma_A-\lambda & 0\\
   \frac{\rho I^*}{\alpha_1 + n^*} - \frac{\rho I^* n^*}{{(\alpha_1 + n^*)}^2} - c_1 I^*& 0 & (\frac{\rho n^*e^{-\lambda \tau}}{\alpha_1 + n^*} - c_1 n^* - \gamma_I)-\lambda
\end{pmatrix} \\
&= det \begin{pmatrix}
    L^*-\lambda & \alpha n^* & -\epsilon n^* \\
    0 & -\gamma_A-\lambda & 0\\
    M^*& 0 & P_1^*+P_2^*e^{-\lambda \tau}-\lambda
\end{pmatrix}
\end{split}
\end{equation*}
where, $L^*$ = $\alpha A^* - \epsilon I^* + r - \frac{2r}{K} n^*$, $M^*$ = $\frac{\rho I^*}{\alpha_1 + n^*} - \frac{\rho I^* n^*}{{(\alpha_1 + n^*)}^2} - c_1 I^*,$  $P_1^*$ = $- c_1 n^* - \gamma_I$ and $P_2^*$ =  $\frac{\rho n^*}{\alpha_1 + n^*}.$
We use the following stability criteria for a nonlinear delay equation from~\cite{kgopal}.
\newtheorem{thm}{Theorem}
\begin{thm} The necessary and sufficient conditions for an equilibrium to be asymptotically stable for all $\tau \geq 0$ are the following:
(1) The real part of the roots of $\Delta(\lambda,0)=0$ is negative.
\\
(2) For all real $\mu$ and all $\tau \geq 0, \ \Delta(i\mu,\tau)\neq 0$.
\end{thm}

\noindent One eigenvalue is of course $-\gamma_A$ and others are given by the root of the following equation
$$(L^*-\lambda) (P_1^*+P_2^*e^{-\lambda \tau}-\lambda)+\epsilon n^*M^*=0.$$ Simplifying, we get

$$\lambda^2+(-L^*-P_1^*-P_2^*e^{-\lambda \tau})\lambda+L^*(P_1^*+P_2^*e^{-\lambda \tau})+\epsilon n^*M^*=0.$$

\noindent We have already obtained condition for non delay case. To examine the other condition, let us substitute $\lambda=i \mu$ in the above equation,
$$-\mu^2+(-L^*-P_1^*-P_2^*e^{-i\mu \tau})i\mu+L^*(P_1^*+P_2^*e^{-i\mu \tau})+\epsilon n^*M^*=0.$$ Separating real and imaginary part, we obtain
\begin{eqnarray}
-\mu^2-P_2^* \sin(\mu \tau)\mu+P_2^*\cos(\mu \tau)+\epsilon n^*M^*&=&0 \nonumber \\
-L^*-P_2^*\cos(\mu \tau)\mu-P_2^*\sin(\mu \tau)&=&0.
\end{eqnarray}
Squaring and adding, we get
$$(\epsilon n^*M^*-\mu^2)^2+{L^*}^2={P_2^*}^2 \mu^2+{P_2^*}^2,$$ which further implies
$$\mu^4-({(P_2^*)}^2+2\epsilon n^*M^*)\mu^2+\epsilon^2 {n^*}^2{M^*}^2+{L^*}^2-{P_2^*}^2=0.$$
The roots are given by
$$\mu^2_{\pm}=\frac{1}{2}\Bigg[{((P_2^*)}^2+2\epsilon n^*M^*))\pm \sqrt{{((P_2^*)}^2+2\epsilon n^*M^*)^2-4(\epsilon^2 {n^*}^2{M^*}^2+{L^*}^2-{P_2^*}^2)}\Bigg]$$
So condition (2) of theorem violates implies that there exists a $\mu$ such that the above relation holds. It holds if
$\epsilon^2 {n^*}^2{M^*}^2+{L^*}^2>{P_2^*}^2,$ so the equilibrium is stable if in addition to the condition for non-delay case (subsection 4.1.2), we have
$\epsilon^2 {n^*}^2{M^*}^2+{L^*}^2<{P_2^*}^2.$

\newtheorem{Remark}{Remark}
\begin{Remark}Stability of Equilibrium with Tea : In this case, since the equation in $I$ does not depend on $T,$ we get the same condition for stability $\epsilon^2 {n^*}^2{M^*}^2+{L^*}^2<{P_2^*}^2$ in addition to the condition given in subsection (4.1.4).
\end{Remark}

\section{Conclusion}
An attempt has been made to model the data obtained from the \textit{in vivo} studies, using Swiss albino mice. Mice are maintained under the influence of iAs for a long period, with iAs-loaded water as the only source of drinking water. Black tea infusion is administered on a regular basis to a subset of the mice. The data obtained on the cell-damage in these two groups of mice were then modelled by the standard dynamical equations. It is redeeming to note that even simple models could fit the data well and predict long-term behaviour of the cell damage by iAs and its alleviation by black tea. Further studies using experimentally obtained fitting parameters would reveal more interesting underlying dynamics that are testable in real systems. The predictions thereof would be reliable and useful for the study of growth and limitation of tumours.

\section*{Appendix}
\noindent We discuss the existence and uniqueness of the solution of the dynamical equations considered in the text above.

\appendix

\section{Existence and uniqueness of the solution}

\noindent We consider the immune-free case (Eq.~\ref{eq7}) first. The case with immunity will follow similarly. The problem can be written as
$$\frac{dy}{dt}=f(t,y), \quad y(0)=y_0, \quad t \in [t_0, t_0+T],$$ where
$f(t,y)=ry(1-\frac{y}{K})+my(1-e^{-\gamma_A t})+pye^{-\gamma_A t}-q y (1-e^{-\gamma_T t})-sye^{-\gamma_T t}$. Since we are interested in only positive solutions, hence we assume the positivity of the solution when $y_0$ is positive. By using bounds of exponentials, we obtain
\begin{eqnarray}
\frac{dy}{dt} & \le & ry(1-\frac{y}{K})+my+py \nonumber \\
& \le & (r+m+p)y -\frac{r}{K}y^2 \nonumber \\
& \le & y((r+m+p) -\frac{r}{K}y).
\end{eqnarray}
Hence, we obtain
$\limsup_{t\rightarrow \infty} y(t) \le \frac{K(r+m+p)}{r}:=M.$ Furthermore, we can write equation as
$$[\frac{1}{y}+\frac{\frac{r}{K}}{r+m+p-\frac{r}{K}y}]dy \le (r+m+p)dt,$$ which gives
$$y(t) \le \frac{r+m+p}{\frac{r}{K}+c e^{-(r+m+p)t}}.$$
Similarly, we get
\begin{eqnarray}
\frac{dy}{dt} & \ge & ry(1-\frac{y}{K})-qy-sy \nonumber \\
& \ge & (r-q-s)y -\frac{r}{K}y^2 \nonumber \\
& \ge & y((r-q-s) -\frac{r}{K}y).
\end{eqnarray}
Hence, we obtain
$\liminf_{t\rightarrow \infty} y(t) \ge \frac{K(r-q-s)}{r}:=m.$
Since all coefficients are positive, hence if $y_0$ is positive, the solution remains positive in this case. So, we have achieved the bounds and positivity of the solution.
In order to ensure the existence and uniqueness of solution, we need to check the Lipschitz condition. Suppose $y_1,y_2 \in \mathbb{R},$ we have
\begin{eqnarray}
&& |f(t,y_1)-f(t,y_2)| \nonumber \\ && \le  r|y_1(1-\frac{y_1}{K})-y_2(1-\frac{y_2}{K})|+m (1-e^{-\gamma_A t}) |y_1-y_2| \nonumber \\ && +p e^{-\gamma_A t}|y_1-y_2|+q (1-e^{-\gamma_T t})|y_1-y_2|+s e^{-\gamma_A t} |y_1-y_2| \nonumber \\
&& \le  r |y_1-y_2|+\frac{r}{K}|y_1-y_2|2M+(m+p+q+s)|y_1-y_2| \nonumber \\
&& \le  (r+m+p+q+s+\frac{2Mr}{K})|y_1-y_2|:=L |y_1-y_2|.
\end{eqnarray}
Hence $f$ is Lipschitz with respect to $y.$ It is easy to see that $f$ is continuous with respect to $t.$ So, we can apply Picard theorem to ensure the existence and uniqueness of solution.

\noindent Next, we discuss the existence and uniqueness of the solution for Eqs.~(\ref{eq8})$-$(\ref{eq11}). One can convert this system of Eqs. in vector from by defining  $X=(n,A,T,I)^t:$
$\frac{dX}{dt}=\hat{F}(X)=AX+F(X),$ where
$$A=
\begin{bmatrix}
r & 0 & 0 & 0 \\
0 & -\gamma_A & 0 &0 \\
0 & 0 & -\gamma_T & 0 \\
0 & 0 & 0 & -\gamma_I
\end{bmatrix}$$
and
$$F(X)=
\begin{bmatrix}
-rn\frac{n}{K} + \alpha A n - \beta n T - \epsilon n I  \\
A_0 \\
T_0  \\
I_0 + \frac{\rho I n}{\alpha_1 + n} -c_1 I n
\end{bmatrix}.$$
It is  straightforward to check that the function $F$ is Lipschitz by assuming the boundedness, i.e. there exists a constant $L$ such that $$\|F(X_1)-F(X_2)\| \leq L_F \|X_1-X_2\|$$
Also, continuity with respect to $t$ is evident since the $F^*$ does not depend on $t.$ Hence, the existence and uniqueness of the solution is guaranteed by Picard-Lindeloff theorem.

\section*{Acknowledgments}
UD would like to acknowledge the Ministry of Human Resource Development (MHRD), India for research fellowship.

\end{document}